\documentclass[12pt,onecolumn,showpacs,floats]{revtex4}
\usepackage[dvips]{graphicx}
\usepackage{dcolumn}
\usepackage{color}

\usepackage{setspace}
\usepackage{bm}
\begin{document}

\title{Is N-doped SrO magnetic? A first-principles view}
\author{Hua Wu}
\thanks{Email address: wuh@fudan.edu.cn}
\affiliation{Laboratory for Computational Physical Sciences, Surface Physics Laboratory, and Department of Physics, Fudan University, Shanghai 200433, China}

\date{today}

\begin{abstract}

N-doped SrO seems to be one of the model systems for $d^0$ magnetism, in which 
magnetism (or ideally, ferromagnetism) was ascribed to the localized N 2$p$ 
spins mediated by delocalized O 2$p$ holes. Here we offer a different view, 
using density functional calculations. We find that N-doped SrO with solely 
substitutional N impurities as widely assumed in the literature is unstable, 
and instead that a pairing state of substitutional and interstitial N impurities
is significantly more stable and has a much lower formation energy than the 
former by 6.7 eV. The stable (N$_{sub}$-N$_{int}$)$^{2-}$ dimers behave like 
a charged (N$_2$)$^{2-}$ molecule and have each a molecular spin=1. 
However, their 
spin-polarized molecular levels lie well inside the wide band gap of SrO and 
thus the exchange interaction is negligibly weak. As a consequence, N-doped SrO
could not be ferromagnetic but paramagnetic.

\end{abstract}


\maketitle

In extensive search of magnetic semiconductors, $d^0$ ferromagnet recently 
draws a lot of attention~\cite{Dietl, Coey, Sanvito,Zunger, Pan, Zhou, Chan, 
Wu, Volnianska, Liu}. It has no magnetic transition-metal or rare-earth element 
present and is free of the issue of clustering of magnetic elements, which, 
however, has been quite often addressed for transition-metal doped 
semiconductors. Thus, an observed $d^0$ ferromagnetism could be intrinsic, 
and it is important both technologically
and in the viewpoint of fundamental physics.

Very recently, a group of density functional calculations~\cite{Kenmochi, 
Elfimov, Shen, Li, Ahuja, Fan, Bluegel09, Kapila, Wang, Bluegel11} 
suggest that when N or C is doped into oxide semiconductors and insulators,
the $2p$ hole carriers introduced by those substitutional impurities 
would be spin-polarized. Due to the dual 
character of the $2p$ holes---a Hund exchange as strong as in $3d$ transition 
metals and delocalization of its orbital~\cite{Elfimov}, 
$d^0$ ferromagnetism could be 
established for the localized N or C 2$p$-hole spins mediated by the 
delocalized O 2$p$ holes. In this respect, those calculations using the 
local-density approximation (LDA) or the generalized gradient approximation 
(GGA) very often give a ferromagnetic and half-metallic band structure. 
In strong contrast, several calculations using the LDA plus Hubbard $U$ method 
~\cite{Pardo} or with a self-interaction correction
~\cite{Droghetti08,Droghetti09}, 
both taking into account a correlation 
effect of localized electrons, show that the $2p$ hole state would be orbitally 
polarized and associated with a Jahn-Teller distortion, thus giving rise to 
an insulating solution with a negligibly weak magnetic coupling. 

In the present Letter, using density functional calculations, we reinvestigate 
N-doped SrO, a recently proposed model system of the $d^0$ 
ferromagnet~\cite{Elfimov}. We will demonstrate that a sole substitution of N 
for O in SrO as widely assumed in the literature is energetically highly 
unfavorable, and instead that a pairing state of substitutional and 
interstitial N dopants is much more stable. The resultant 
(N$_{sub}$-N$_{int}$)$^{2-}$ dimers carry each a molecular spin=1, 
and their spin-polarized levels lie well inside the wide band gap of SrO. 
As a result, N-doped SrO seems not to be a ferromagnetic half-metal 
with an atomic 
spin as often predicted in the literature, but a paramagnetic insulator with
a molecular spin. Thus, this work suggests that much care needs to be taken 
when categorizing N-doped oxide semiconductors and insulators into the 
$d^0$ ferromagnet.

We performed LDA calculations, using the full-potential augmented plane
wave plus local orbital code (WIEN2k)~\cite{Blaha}. 
The experimental lattice constant of 5.14 \AA~
was used for SrO. A 2$\times$2$\times$2 supercell was introduced to 
simulate the substitutional and interstitial N impurities. 
The muffin-tin spheres were chosen to be 2.5~Bohr for Sr and 1.1 Bohr 
for O and N, the plane-wave cutoff 
of 12~Ry for the interstitial wave functions, and a 5$\times$5$\times$5 
{\bf k}-point mesh for integration over the Brillouin zone. 
Structural relaxations were carried out till the atomic force is less than 
50 meV/\AA~. 
Note that our calculations using GGA gave 
almost the same results as LDA. 

We first calculate two isolated N$_{sub}$ impurities in the 
Sr$_{32}$O$_{30}$N$_2$ supercell having the N-N distance 
of 8.9 \AA~. This distance is so long that this structure also mimics single 
N$_{sub}$ impurities. The calculation gives a half-metallic solution as 
reported in the literature. The obtained density of states (DOS) is shown in 
Fig. 1. This is because each N$_{sub}$ impurity introduces one 2$p$ hole and 
the strong Hund exchange makes the N 2$p$ majority-spin orbitals fully occupied 
and the minority-spin orbitals 2/3 filled. The sharp DOS peak at the Fermi 
level suggests that this structure could be unstable. The LDA calculations, 
due to a self-interaction error of localized electrons, overestimate electron 
delocalization, and thus a (long-range) ferromagnetic exchange interaction 
between the N$_{sub}$ impurities via the O 2$p$ is highly 
overestimated~\cite{Pardo, Droghetti08, Droghetti09}. 

\begin{figure}[t]
\centering \includegraphics[width=10cm]{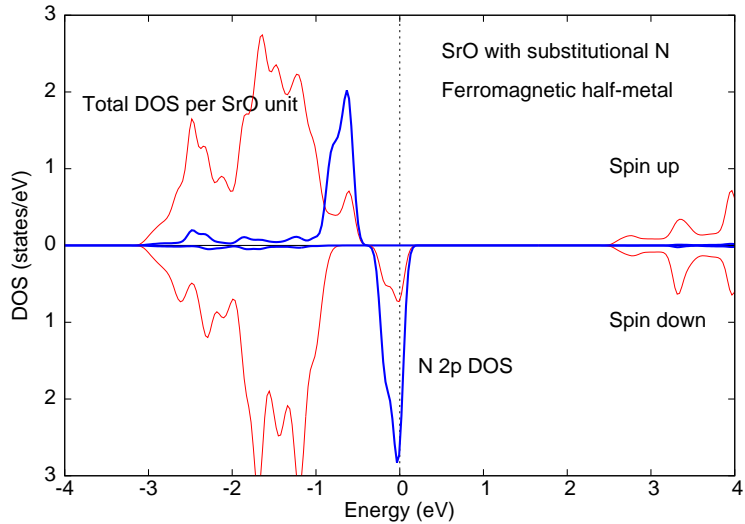} \caption{
(Color online) \label{SrO_2Nsub}
Total DOS (red thin curves) and N $2p$ DOS (blue bold curves) calculated 
by LDA for a 
Sr$_{32}$O$_{30}$N$_2$ (2$\times$2$\times$2) supercell having two isolated 
substitutional N impurities. The upper (lower) panel shows the spin up (down)
channel. Fermi level is set at zero energy. It is calculated to be 
a ferromagnetic half-metal having an atomic spin=1/2 carried mainly by each
N$^{2-}$ impurity.}
\end{figure}

As N atom has a high electronegativity and the 2$p$ hole state of the 
N$_{sub}$ impurities in SrO would be unstable, the N$_{sub}$ impurities 
could prefer a pairing. We therefore calculate the pairing state, assuming 
two N$_{sub}$ impurities sitting on two nearest neighboring O sites 
(with the distance of about 3.6 \AA~). After a structural relaxation, 
the pairing state is found to have a lower energy than the above isolated 
impurity 
state by 170 meV and the two N$_{sub}$ impurities move closer by only 
0.12 \AA~ (each N$_{sub}$ impurity displaces from the ideal O vacancy site by 
0.06 \AA~). Obviously, this is not a true pairing state, in view of the 
small energy gain (in terms of a chemical bonding energy) and the long 
N-N distance (of 3.5 \AA). The obtained solution is 
still a ferromagnetic half-metal with spin=1/2 for each N$_{sub}$ impurity 
(not shown here). If that were a true pairing state, the solution would be 
a nonmagnetic insulator having closed molecular orbitals of the 
(N$_{sub}$-N$_{sub}$)$^{4-}$ dimers 
(cf. Fig. 2(c), with two more electrons filling up the $pp\pi^*$ antibonding
state, and leaving only the highest-level $pp\sigma^*$ antibonding state 
unoccupied). But if the N$_{sub}$ impurities formed a dimer and thus strongly 
displaced from 
the ideal O vacancy sites, there would be a huge energy cost associated with
the large lattice deformation. This would prevent the 
N$_{sub}$ impurities forming the dimers. 

So far, the stability of the solely substitutional N impurities in SrO 
has been a remaining issue. As SrO has a large lattice spacing, 
interstitial N impurities 
(N$_{int}$) are also possible, in addition to the widely assumed 
N$_{sub}$ impurities. As a result, a N$_{sub}$-N$_{int}$ pairing state 
could be a right way to enhance the stability of N-doped SrO. 
This is indeed the case, as supported by the following calculations.

\begin{figure}[t]
\centering \includegraphics[width=10cm]{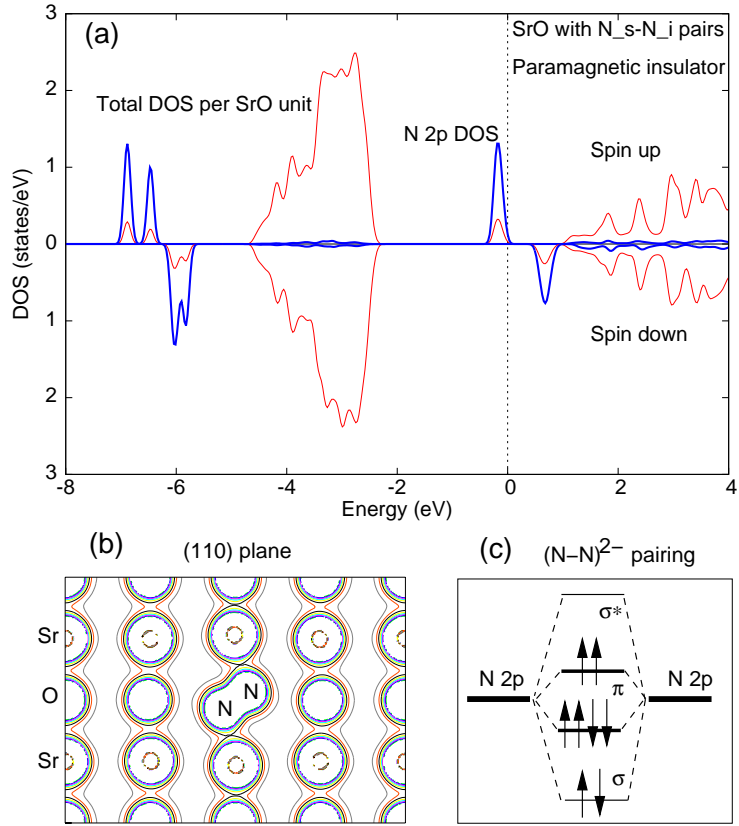} \caption{
(Color online) \label{SrO_Nsub_Nint}
(a) Total DOS (red thin curves) and N $2p$ DOS (blue bold curves) calculated 
by LDA for a Sr$_{32}$O$_{31}$N$_2$ 
(2$\times$2$\times$2) supercell having a pair of substitutional and interstitial
N impurities. The upper (lower) panel shows the spin up (down) channel. 
Fermi level is set at zero energy. After a full atomic relaxation, the pair 
of N impurities becomes a (N$_2$)$^{2-}$ dimer, which symmetrically straddles 
the O vacancy site along a body diagonal of the cubic supercell, see (b) a charge
density contour plot (0.1-0.8 $e$/\AA$^3$) on the (110) plane. (c) A schematic
energy level diagram due to the (N-N)$^{2-}$ pairing interaction. The
antibonding $pp\pi^*$ doublet state is half filled and carries a molecular
spin=1. This spin-split $pp\pi^*$ state lies inside the wide band gap of 
SrO [see (a)], giving rise to a paramagnetic insulating state.}
\end{figure}

We then calculate the Sr$_{32}$O$_{31}$N$_2$ supercell having one N$_{sub}$ 
impurity and one N$_{int}$ impurity nearby. After a full structural 
relaxation, 
the N$_{sub}$ impurity displaces from the ideal O vacancy site, and 
the initialized 
N$_{sub}$-N$_{int}$ pair becomes a (N$_2$)$^{2-}$ dimer, which 
symmetrically straddles the 
O vacancy site along a body diagonal of the cubic supercell, 
see Fig. 2(b). 
The resultant dimer has a short bondlength of 1.25 \AA~ and thus gains 
a lot of energy due to the strong chemical bonding. Taking the free 
O$_2$ molecules as a reservoir, we compare the total energy of the 
supercells Sr$_{32}$O$_{30}$N$_2$ (plus 1/2 O$_2$) and Sr$_{32}$O$_{31}$N$_2$. 
We find that the latter having the (N$_2$)$^{2-}$ dimer has a much lower 
formation energy than the former by 6.7 eV. Thus the stability of N-doped 
SrO is indeed significantly enhanced via the dimerization of the 
N$_{sub}$-N$_{int}$ impurities, compared to the sole N$_{sub}$ impurities 
widely assumed in the literature. The molecular level diagram of the 
(N$_2$)$^{2-}$ dimer is schematically plotted in Fig. 2(c). 
The $pp\pi^*$ antibonding state is half filled and thus the (N$_2$)$^{2-}$ dimer 
has a molecular spin=1 like the gas-phase O$_2$ molecules. 
A similar molecular magnetism was studied in our previous work~\cite{Wu}
and addressed in a recent review article~\cite{Volnianska}. Note, however, 
that the spin-split $pp\pi^*$ level lies well inside the wide band gap of SrO 
(albeit the experimental gap of 5.3 eV is underestimated to be 3.5 eV 
by LDA), see the insulating solution shown in Fig. 2(a). As a consequence, 
the exchange interaction between those magnetic dimers, if any, 
would be superexchange like and tiny. This is indeed supported by our 
calculations which show that an antiferromagnetic state has a `lower' 
energy than a ferromagnetic state by only 1 meV. 
Those calculations were performed for a Sr$_{32}$O$_{30}$N$_4$ supercell,
which has two (N$_2$)$^{2-}$ dimers symmetrically straddling 
the corner and body-center oxygen vacancy sites, respectively, 
along a body diagonal of the 2$\times$2$\times$2 supercell.
The inter-dimer distance (from one center to the other center) is 8.9 \AA.
The intra-dimer N-N distance remains to be 1.25 \AA~ after a full 
atomic relaxation.
Taking the computational accuracy into account,  
we could propose that practically there is no magnetic coupling between 
those magnetic dimers. Thus, N-doped SrO would be a paramagnetic insulator.

To summarize, using LDA electronic structure calculations, we cast a doubt 
on the suitability of handling N-doped SrO as a model system of the 
$d^0$ ferromagnet. Our calculations show that the solely substitutional 
N impurities in SrO, widely assumed in the literature, are unstable. 
The corresponding ferromagnetic half-metallic solution with atomic $2p$ hole
states could thus be an artifact. 
Instead, we find that a N$_{sub}$-N$_{int}$ pairing state, 
with (N$_2$)$^{2-}$ dimers each straddling an oxygen vacancy site symmetrically, 
is much more stable.
Although those (N$_2$)$^{2-}$ dimers have each a molecular spin=1, 
their magnetically active levels lie well inside the wide band gap of SrO 
and thus the superexchange interaction between those magnetic dimers is 
negligibly weak. As a consequence, N-doped SrO could not a ferromagnetic 
half-metal with an atomic spin, but a paramagnetic insulator with a molecular
spin. As the atomic 2$p$ hole state has quite often 
been used to predict the $d^0$ ferromagnetism in the N or C doped oxide 
semiconductors and insulators, the present study provides a warning message 
against such predictions.
%
%

.


\begin{thebibliography}{100}

\bibitem{Dietl} T.~Dietl, Nat. Mater. \textbf{9}, 965 (2010). 

\bibitem{Coey} J. M. D. Coey, Solid State Sci. {\bf 7}, 660 (2005).

\bibitem{Sanvito} C. D. Pemmaraju and S. Sanvito, 
Phys. Rev. Lett. {\bf 94}, 217205 (2005). 

\bibitem{Zunger} A. Zunger, S. Lany, and H. Raebiger, 
Physics {\bf 3}, 53 (2010).

\bibitem{Pan} H. Pan, J. B. Yi, L. Shen, R. Q. Wu, J. H. Yang, J. Y. Lin, 
Y. P. Feng, J. Ding, L. H. Van, and J. H. Yin, 
Phys. Rev. Lett. {\bf 99}, 127201 (2007).

\bibitem{Zhou}
S. Zhou, Q. Xu, K. Potzger, G. Talut, R. Gr\"otzschel, J. Fassbender, 
M. Vinnichenko, J. Grenzer, M. Helm, H. Hochmuth, M. Lorenz, M. Grundmann, 
and H. Schmidt, Appl. Phys. Lett. {\bf 93}, 232507 (2008).

\bibitem{Chan} J. A. Chan, S. Lany, and A. Zunger,
Phys. Rev. Lett. {\bf 103}, 016404 (2009).

\bibitem{Wu} H. Wu, A. Stroppa, S. Sakong, S. Picozzi, M. Scheffler, and 
P. Kratzer, Phys. Rev. Lett. {\bf 105}, 267203 (2010).

\bibitem{Volnianska} O. Volnianska and P. Boguslawski, 
J. Phys.: Condens. Matter {\bf 22}, 073202 (2010).

\bibitem{Liu} Yu Liu, Gang Wang, Shunchong Wang, Jianhui Yang, Liang Chen, 
Xiubo Qin, Bo Song, Baoyi Wang, and Xiaolong Chen,
Phys. Rev. Lett. {\bf 106}, 087205 (2011).

\bibitem{Kenmochi} K. Kenmochi, M. Seike, K. Sato, A. Yanase, 
and H. Katayama-Yoshida, Jpn. J. Appl. Phys. {\bf 43}, L934 (2004). 

\bibitem{Elfimov} I. S. Elfimov, A. Rusydi, S. I. Csiszar, Z. Hu, 
H. H. Hsieh, H.-J. Lin, C. T. Chen, R. Liang, and G. A. Sawatzky, 
Phys. Rev. Lett. {\bf 98}, 137202 (2007).

\bibitem{Shen} L. Shen, R. Q. Wu, H. Pan, G. W. Peng, M. Yang, Z. D. Sha, 
and Y. P. Feng, Phys. Rev. B {\bf 78}, 073306 (2008).

\bibitem{Li} Haowei Peng, H. J. Xiang, S.-H. Wei, Shu-Shen Li, 
Jian-Bai Xia, and Jingbo Li, Phys. Rev. Lett. {\bf 102}, 017201 (2009).

\bibitem{Ahuja} Xiangyang Peng and Rajeev Ahuja,
Appl. Phys. Lett. {\bf 94}, 102504 (2009).

\bibitem{Fan} S. W. Fan, K. L. Yao, and Z. L. Liu,
Appl. Phys. Lett. {\bf 94}, 152506 (2009).

\bibitem{Bluegel09} P. Mavropoulos, M. Le\v{z}ai\'c, and S. Bl\"ugel,
Phys. Rev. B {\bf 80}, 184403 (2009).

\bibitem{Kapila} N. Kapila, V. K. Jindal, and H. Sharma,
J. Phys.: Condens. Matter {\bf 23}, 446006 (2011).

\bibitem{Wang} Q. J. Wang, J. B. Wang, X. L. Zhong, Q. H. Tan, 
and Y. C. Zhou, Europhys. Lett. {\bf 95}, 47010 (2011). 

\bibitem{Bluegel11} I. Slipukhina, P. Mavropoulos, S. Bl\"ugel, 
and M. Le\v{z}ai\'c, Phys. Rev. Lett. {\bf 107}, 137203 (2011). 

\bibitem{Pardo} V. Pardo and W. E. Pickett, 
Phys. Rev. B {\bf 78}, 134427 (2008).

\bibitem{Droghetti08} A. Droghetti, C. D. Pemmaraju, and S. Sanvito,
Phys. Rev. B {\bf 78}, 140404(R) (2008).

\bibitem{Droghetti09} A. Droghetti and S. Sanvito,
Appl. Phys. Lett. {\bf 94}, 252505 (2009).

\bibitem{Blaha} P. Blaha, K. Schwarz, G. Madsen, D. Kvasnicka, and J. Luitz,
{\bf WIEN2k}, 2001. ISBN 3-9501031-1-2.

\end{thebibliography}
\end{document}